\documentclass[prl,aps, 
twocolumn, 
showpacs,
superscriptaddress,
10pt,nofootinbib]{revtex4-1} 

\usepackage[T1]{fontenc}
\usepackage{bm,dcolumn}
\usepackage{amsmath,amsfonts,amssymb,amsthm,amscd} 
\usepackage{graphicx}
\usepackage{verbatim}
\usepackage[section]{placeins}
\usepackage{color}
\usepackage{soul}
\usepackage{xparse}
\usepackage{physics}
\usepackage{siunitx}
\usepackage{CJK}

\graphicspath{ {./Figures/} }
\usepackage{pdfpages} 
\usepackage{pgffor}

\makeatletter
\AtBeginDocument{\let\LS@rot\@undefined}
\makeatother

\begin{document}
%\begin{CJK*}{GB}{} 

\title{Dark state optical lattice with sub-wavelength spatial structure}

\author{Y. Wang}
\email{yuw127@umd.edu}
\affiliation{Joint Quantum Institute, National Institute of Standards and Technology and the University of Maryland, College Park, Maryland 20742 USA}

\author{S. Subhankar}
\affiliation{Joint Quantum Institute, National Institute of Standards and Technology and the University of Maryland, College Park, Maryland 20742 USA}

\author{P. Bienias}
\thanks{These two authors contributed equally}
\affiliation{Joint Quantum Institute, National Institute of Standards and Technology and the University of Maryland, College Park, Maryland 20742 USA}

\author{M. \L\k{a}cki}
\thanks{These two authors contributed equally}
\affiliation{Jagiellonian University, Institute of Physics, \L ojasiewicza 11, 30-348 Krak\'ow, Poland}
\affiliation{Institute for Quantum Optics and Quantum Information of the Austrian Academy of Sciences, A-6020 Innsbruck, Austria}
\affiliation{Institute for Theoretical Physics, University of Innsbruck, A-6020 Innsbruck, Austria}

\author{T-C. Tsui}
\affiliation{Joint Quantum Institute, National Institute of Standards and Technology and the University of Maryland, College Park, Maryland 20742 USA}

\author{M. A. Baranov}
\affiliation{Institute for Quantum Optics and Quantum Information of the Austrian Academy of Sciences, A-6020 Innsbruck, Austria}
\affiliation{Institute for Theoretical Physics, University of Innsbruck, A-6020 Innsbruck, Austria}

\author{A. V. Gorshkov}
\affiliation{Joint Quantum Institute, National Institute of Standards and Technology and the University of Maryland, College Park, Maryland 20742 USA}
\affiliation{Joint Center for Quantum Information and Computer Science, National Institute of Standards and Technology and the University of Maryland, College Park, Maryland 20742 USA}

\author{P. Zoller}
\affiliation{Institute for Quantum Optics and Quantum Information of the Austrian Academy of Sciences, A-6020 Innsbruck, Austria}
\affiliation{Institute for Theoretical Physics, University of Innsbruck, A-6020 Innsbruck, Austria}

\author{J. V. Porto}
\affiliation{Joint Quantum Institute, National Institute of Standards and Technology and the University of Maryland, College Park, Maryland 20742 USA}

\author{S. L. Rolston}
\affiliation{Joint Quantum Institute, National Institute of Standards and Technology and the University of Maryland, College Park, Maryland 20742 USA}

\date{\today}

\begin{abstract}

We report on the experimental realization of a conservative optical lattice for cold atoms with sub-wavelength spatial structure. The potential is based on the nonlinear optical response of three-level atoms in laser-dressed dark states, which is not constrained by the diffraction limit of the light generating the potential. The lattice consists of a 1D array of ultra-narrow barriers with widths less than 10~nm, well below the wavelength of the lattice light, physically realizing a Kronig-Penney potential. We study the band structure and dissipation of this lattice, and find good agreement with theoretical predictions. The observed lifetimes of atoms trapped in the lattice are as long as 60 ms, nearly $10^5$ times the excited state lifetime, and could be further improved with more laser intensity.  The potential is readily generalizable to higher dimension and different geometries, allowing, for example, nearly perfect box traps, narrow tunnel junctions for atomtronics applications, and dynamically generated lattices with sub-wavelength spacings.
\end{abstract}


\maketitle
%\end{CJK*}

Coherent control of position and motion of atoms with laser light
has been a primary enabling technology in the physics of ultracold atoms.
The paradigmatic examples of conservative optical potentials are the
optical dipole trap and optical lattices, generated by far off-resonant
laser fields, with the ac-Stark shift of atomic levels
as the underlying mechanism. The scale and spatial resolution
for such optical potential landscapes is determined by the diffraction
limit, which is of order the wavelength of the light $\lambda$. 
This fundamentally limits optical manipulation of atoms. 
For example, in quantum simulation with cold atoms in optical
lattices, the minimum lattice constant is $\lambda/2$, setting
the energy scale for Hubbard models for both hopping (kinetic
energy) and interaction of atoms, with challenging temperature
requirements to observe quantum phases of interest (see~\cite{Gross2017} and references therein). Developing tools
to overcome the diffraction limit, allowing coherent
optical manipulation of atoms on the sub-wavelength scale, is thus
an outstanding challenge. Following recent theoretical proposals~\cite{Dum1996,Lacki2016,Jendrzejewski2016}
we report below first experiments demonstrating coherent optical potentials with
sub-wavelength spatial structure by realizing a Kronig-Penney type optical lattice with barrier widths below $\lambda/50$.

In the quest to beat the diffraction limit, several ideas have been proposed to create coherent optical potentials with sub-wavelength structure. 
These include Fourier-synthesis of lattices using multiphoton Raman transitions~\cite{Ritt2006,Salger2007}, optical or radio-frequency dressing of optical potentials~\cite{yi08a, lundblad08a}, and trapping in near-field guided modes with nano-photonic systems~\cite{GonzalezTudela2015,Gullans2012} (although they suffer from decoherence induced by nearby surfaces). An alternative approach uses the spatial dependence of the nonlinear atomic response associated with the dark state of a three-level system~\cite{Gorshkov2008,Kiffner2008,Miles2014, Sahrai2005, Cho2007, Kapale2006}, as a means to realize sub-wavelength atomic addressing and excitation. The sub-wavelength resolution arises when optical fields are arranged so that the internal dark state composition varies rapidly (``twists'') over a short length scale. 

As proposed in~\cite{Lacki2016,Jendrzejewski2016}, such a sub-wavelength twist can also be used to create a conservative potential with narrow spatial extent, due to the energy cost of the kinetic energy term of the Hamiltonian~\cite{Dum1996,Dutta1999,Cheneau2008}. Unlike ac-Stark shift potentials, this twist-induced potential is a quantum effect, with magnitude proportional to $\hbar$.
Using this effect, we create 1D lattices with barrier widths less than $\lambda/50$, where $\lambda$ is the wavelength of lattice light. This potential realizes the Kronig-Penney (KP) lattice model~\cite{Kronig499}---a lattice of nearly $\delta$-function potentials. We study the band structure and dissipation, and find that the dark state nature of this potential results in suppressed scattering, in good agreement with theoretical models. 
%TODO despite on resonant 

%-------------------------------------------FIGURE 1------------------------------------------------------------------
\begin{figure}[!t]
\includegraphics[width=8.cm]{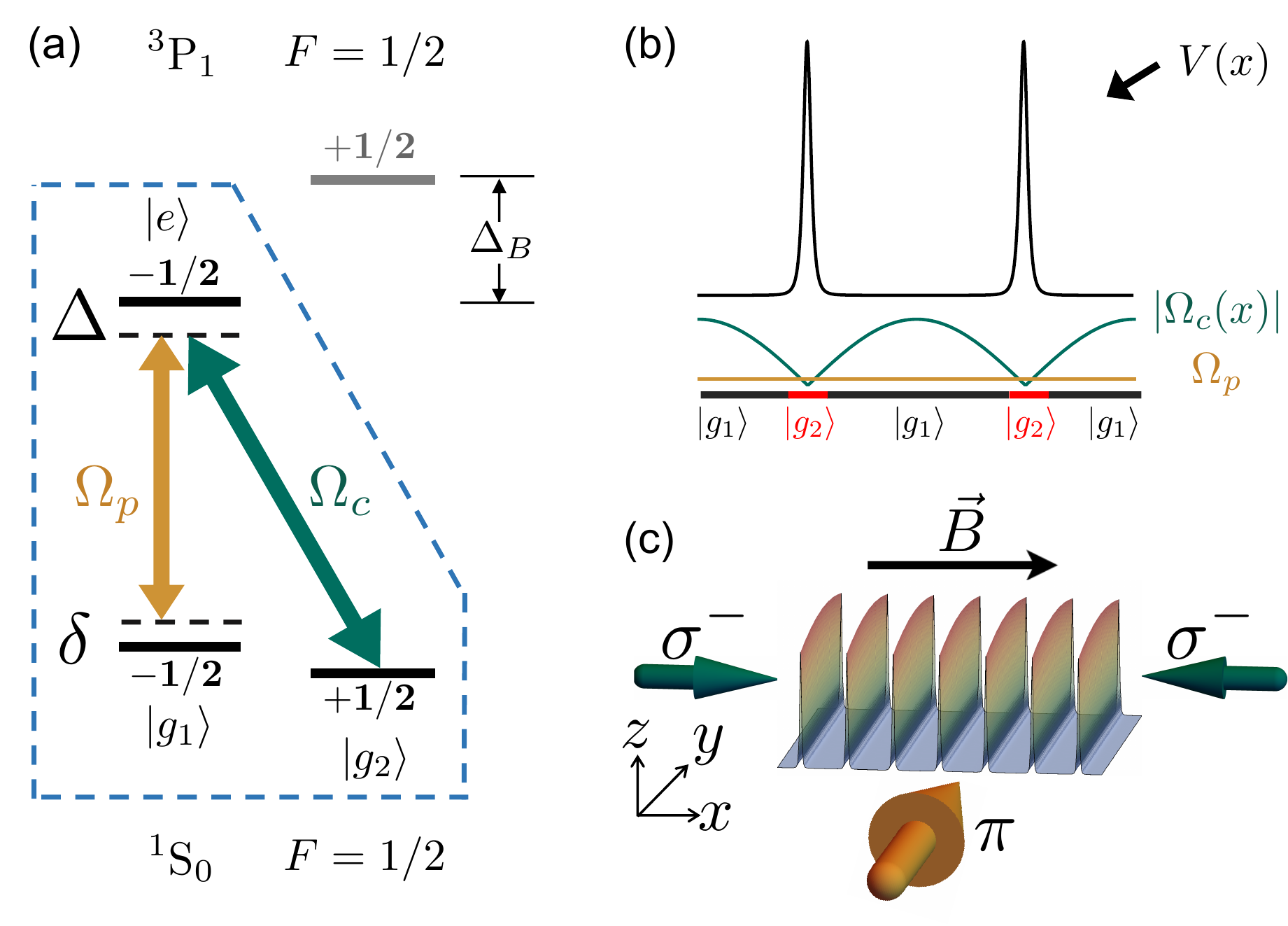}
\caption{(Color online) Level structures and experimental geometry.
\textbf{(a)} The three levels in $^{171}$Yb used to realize the dark state are isolated from the fourth $^3$P$_1$, $m_F=+1/2$ state by a large magnetic field. They are coupled by a strong $\sigma^-$ polarized control field $\Omega_c$ (green) and a weak $\pi$ polarized probe field $\Omega_p$ (orange). The resulting dark state is a superposition of the ground states $\ket{g_1}$ and $\ket{g_2}$, with relative amplitudes determined by $ \Omega_c(x)/\Omega_p$.
\textbf{(b)} Spatial dependence of the dark state composition is created using a standing wave control field $\Omega_c(x)$, and a traveling wave probe field $\Omega_p$. The geometric potential $V(x)$ (black) arises as the dark state rapidly changes its composition near the nodes of the standing wave.
\textbf{(c)} The two counter-propagating $\sigma^-$ beams creating the standing wave are aligned with a strong magnetic field along $x$, while the $\pi$ beam travels along $y$.} 
\label{fig:schematic}
\end{figure}
%%-------------------------------------------FIGURE 1-------------------------------------------------------------------

Our approach is illustrated in Fig.~\ref{fig:schematic}~(a). A three-level system is coupled in a $\Lambda$-configuration by two optical fields: a spatially varying strong control field $\Omega_c(x)= \Omega_c \sin{(k x)}$ and a constant weak probe field $\Omega_p$. The excited state $|e\rangle$ can decay to either ground state $|g_i\rangle$. Within the Born-Oppenheimer (BO) approximation, slowly-moving atoms in the dark state $|E_0(x)\rangle$ are decoupled from $|e\rangle$, where $|E_0(x)\rangle = \sin(\alpha) |g_1\rangle-\cos(\alpha) |g_2\rangle$, and $\alpha(x) = \arctan[\Omega_c(x)/\Omega_p]$~\cite{Lacki2016}. The two bright states $E_{\pm}(x)$ have excited state component $|e\rangle$, leading to light scattering. As shown in Fig.~\ref{fig:schematic}~(b), the fields are arranged in such a way that the dark state changes composition over a narrow region in space, depending on the ratio $\epsilon= \Omega_p / \Omega_ c$. The kinetic energy associated with this large gradient in the spin wavefunction gives rise to a conservative optical potential $V(x)$~\cite{Lacki2016,Jendrzejewski2016} for atoms in $|E_0(x)\rangle$,

\begin{equation}
V(x)=\frac{\hbar^2}{2m} \left(\frac{d\alpha}{dx}\right)^2  =E_R\frac{\epsilon^2 \textrm{cos}^2(kx)}{[\epsilon^2+\textrm{sin}^2(kx)]^2}
\end{equation}
where $k=2\pi/\lambda$, $E_R=\hbar^2 k^2/2m$ is the recoil energy, $m$ is the mass of the atom. The potential $V(x)$ can be viewed as arising from non-adiabatic corrections to the BO potential~\cite{Lacki2016,Jendrzejewski2016} or artificial scalar gauge potential~\cite{Cheneau2008,Dalibard2011,Goldman2014}. 
When $\epsilon\ll 1$ this creates a lattice of narrow barriers spaced by $\lambda/2$, with the barrier height scaling as $1/\epsilon^2$, and the full width half maximum scaling as $0.2 \lambda \epsilon$ (Fig.~\ref{fig:schematic}(b)).

The potential $V(x)$ exhibits several properties that distinguish it from typical optical potentials based on ac-Stark shifts: (1) the explicit dependence on $\hbar$, via the recoil energy $E_R$, reveals the quantum nature of $V(x)$ arising from the gradient in the atomic wavefunction, whereas a typical optical potential can be described entirely classically as an induced dipole interacting with the electric field of the laser; (2) since gradients in wavefunctions always cost energy, $V(x)$ is always repulsive; (3) the geometric nature of the potential results in it being only dependent on $\epsilon$. By deriving both fields from the same laser it is relatively insensitive to technical noise; and (4) unlike near-field guided modes~\cite{GonzalezTudela2015,Gullans2012}, our scheme works in the far field, thus avoiding the decoherence associated with the proximity of surfaces.

We realize the $\Lambda$-configuration using three states selected from the  $^1$S$_0$, $F=1/2$ and $^3$P$_1$, $F=1/2$ hyperfine manifolds in $^{171}$Yb. The two $^1$S$_0$ ground states $m_F=\pm1/2$ comprise the lower two states $|g_1\rangle$ and $|g_2\rangle$ (see Fig.~\ref{fig:schematic}~(a)). The $^3$P$_1$, $m_F=-1/2$ state, with inverse lifetime $\Gamma=2\pi\times182$ kHz, makes up the third state $|e\rangle$ in the $\Lambda$-configuration. The $|g_i\rangle \rightarrow |e\rangle$ transitions are isolated from transitions to the other $^3$P$_1$, $m_F=+1/2$ states by applying a 12 mT magnetic field $\vec{\bm{B}}$ to Zeeman split the two $^3$P$_1$ states by 
$\Delta_B = 1.8\times 10^3 ~\Gamma$. 
The same field only slightly splits the $^1$S$_0$ ground states by -0.5 $\Gamma$ due to the small nuclear magnetic moment.  
The standing-wave control field $\Omega_c(x)$, traveling along $\vec{\bm{B}}$, is produced by two counter-propagating $\sigma^-$ polarized laser beams that couple the $|g_2\rangle$ and $|e\rangle$ states with independently controlled amplitudes $\Omega_{c1} e^{i k x}$ and $\Omega_{c2} e^{-i k x}$. A third beam, $\pi$ polarized and traveling normal to $\vec{\bm{B}}$, couples the $|g_1\rangle$ and $|e\rangle$ states with amplitude $\Omega_p e^{i k y}$.  The laser frequency of the control and probe beams can be chosen to set the single-photon and two-photon detunings, $\Delta$ and $\delta$. We define $\delta=0$ as the dark state condition for the isolated three-level system, accounting for the Zeeman splitting. Off-resonant couplings to other states can introduce light shifts which require nonzero $\delta$ to maintain the dark state condition.

We create an ultracold $^{171}$Yb gas by sympathetically cooling it with Rb atoms in a bichromatic crossed dipole trap~\cite{Vaidya2015, Herold2012}. After Yb atoms are collected in the trap with a temperature of $\simeq300$ nK ($T/T_F= 1.10$, where $T_F$ is the Fermi temperature), the magnetic field in the $x$ direction is ramped up in 100~ms to 12 mT, removing Rb atoms from the trap. The Yb atoms are then optically pumped into $|g_1\rangle$ using a 50~ms pulse from one of the control beams, resulting in $\simeq 1.5\times10^5$ Yb atoms polarized. The small $^{171}$Yb scattering length (-3$a_0$,  with $a_0$ the Bohr radius), plus the lack of $s$-wave scattering in polarized fermions allow us to safely neglect interactions. The Rabi frequencies of each of the three beams are calibrated by measuring the two-photon Rabi frequencies from $|g_1\rangle \rightarrow |g_2\rangle$ at large $\Delta$ with different pairs of beams. The laser polarization purity and alignment to $\vec{\bm{B}}$ are carefully optimized, such that the residual fraction of wrong polarization measured in Rabi frequency is less than 0.5\%. To load Yb atoms into the ground band of the dark state lattice, we first populate the spatially homogeneous dark state by ramping on $\Omega_{c1}$ followed by $\Omega_p$, and then adiabatically ramp on $\Omega_{c2}$ in 1 ms, creating the lattice. We measure the momentum distribution using a band mapping sequence~\cite{Kastberg1995}, by first ramping off $\Omega_{c2}$ in 0.5 ms, and then suddenly turning off all the other light fields. We then take absorption images after time-of-flight (TOF)  along $y$ to measure the momentum along $x$ and $z$. See~\cite{supplementary} for further details.

%-------------------------------------------FIGURE 2------------------------------------------------------------------
\begin{figure}[!t]
\includegraphics[width=8cm]{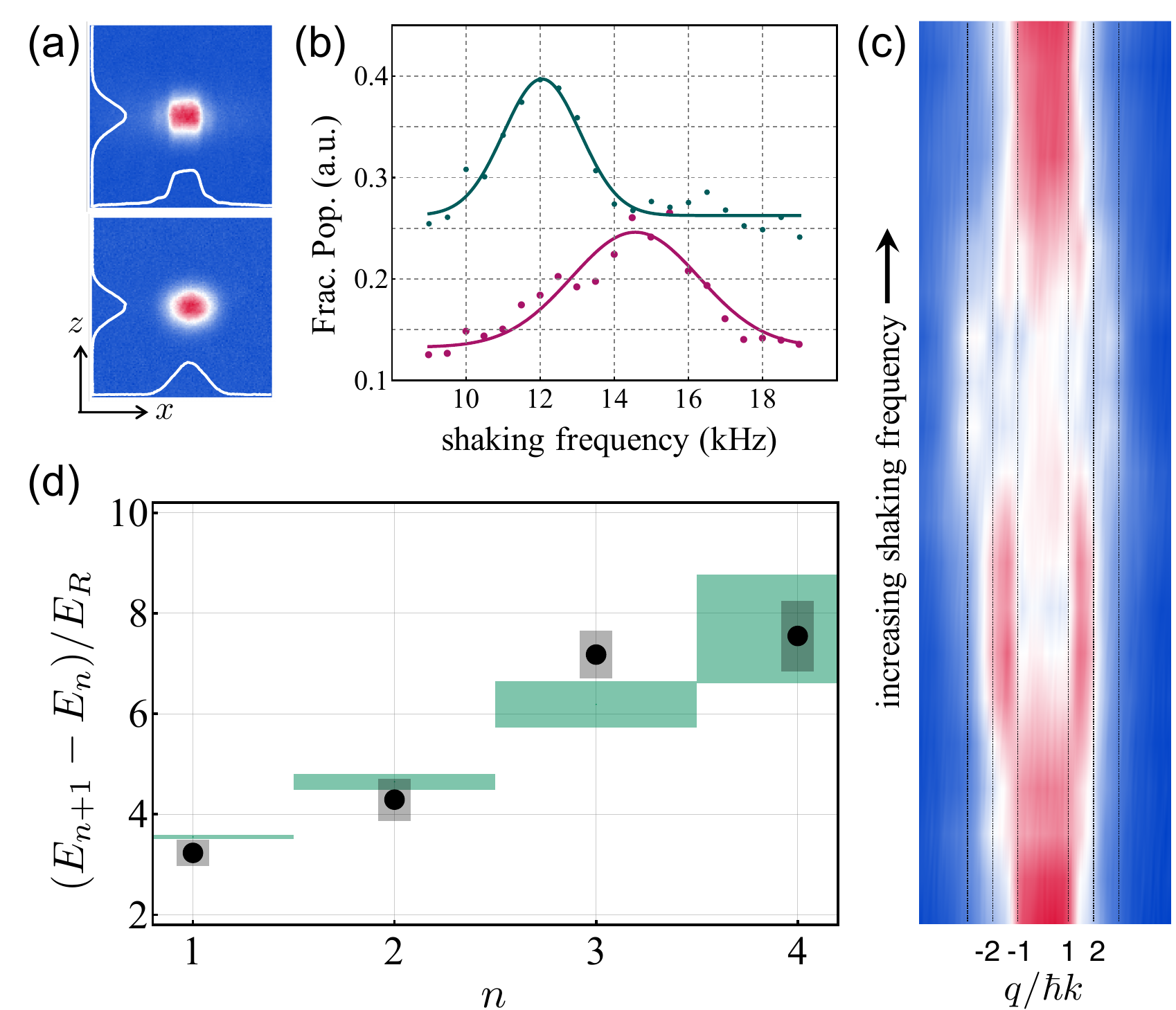}
\caption{(Color online) 
\textbf{(a)} Band mapping results for atoms loaded into the dark state lattice with three beams (upper), and with only $\Omega_c$ beams (lower). 
The white traces show the integrated momentum distribution in each direction  ($x$ is the lattice direction). 
\textbf{(b,c)} Band spectroscopy: 
in (c) we plot the TOF column density integrated over $z$ after shaking the lattice vs. the shaking frequency; in (b) we plot the fraction of the population excited to the $p$-band (dark green) and $d$-band (magenta) Brillouin zones (see (c)) vs. shaking frequency. Gaussian fits (colored lines in (b)) are used to determine the center frequency and the width of the transition. 
\textbf{(d)} Band spacing scaling: $E_{n+1}-E_n$ is plotted vs. the band index $n$ of a dark state lattice with $\Omega_c=70\Gamma$, $\Omega_p=10\Gamma$, $\Delta=22\Gamma$ and $\delta=0$. The grey vertical bars indicate the transition width inferred from the measurements, while the green rectangles are predictions of the expected band spacings and widths~\cite{supplementary}. 
}
\label{fig:shake}
\end{figure}
%-------------------------------------------FIGURE 2-------------------------------------------------------------------

The existence of lattice structure of $V(x)$ leads to Brillouin zones (BZ), visible in TOF images taken after band mapping. Since $k_{\textrm{B}} T$ is less than the band gap, the population is predominantly in the first BZ and distinct band edges are visible (upper panel in Fig.~\ref{fig:shake}~(a)). The lower panel shows the result with no probe beam, where we find a nearly Gaussian distribution in the lattice direction. We also see nearly Gaussian distributions for atoms loaded in the other two-beam configurations: $\Omega_{c1}\ \&\ \Omega_p$ and $\Omega_{c2}\ \&\ \Omega_p$.

For small $\epsilon$, this lattice maps to a 1D KP model. One characteristic feature of the KP lattice is that the energy of the $n$th-band scales as $n^2E_R$, such that the band spacing {\em{increases}} with $n$. In contrast, in a deep sinusoidal lattice the band spacing {\em{decreases}} with $n$. To map out the band structure, we excite atoms from the ground ($s$-) band into the higher bands by shaking the lattice using phase modulation of one of the $\sigma^-$ beams. After band mapping we measure the band populations, which become spatially separated after TOF (see Fig.~\ref{fig:shake}~(c)). Fig.~\ref{fig:shake}~(b) plots the frequency-dependent excitation into the first ($p$-) and second ($d$-) excited bands for $\epsilon=0.14$, extracted from the data in Fig.~\ref{fig:shake}~(c). The $s\rightarrow d$ excitation arises from a two-step process involving the intermediate $p$-band. We map out the band structure up to the $g$-band and plot the energy differences for adjacent bands (see Fig.~\ref{fig:shake}~(d)), which increases monotonically with $n$. 
The green rectangles show the theoretical band spacings and widths, calculated from a model that includes both the light shifts from states outside the three-level system~\cite{supplementary}, and mixing with the bright states. 

%-------------------------------------------FIGURE 3------------------------------------------------------------------
\begin{figure}[!t]
\includegraphics[width=8cm]{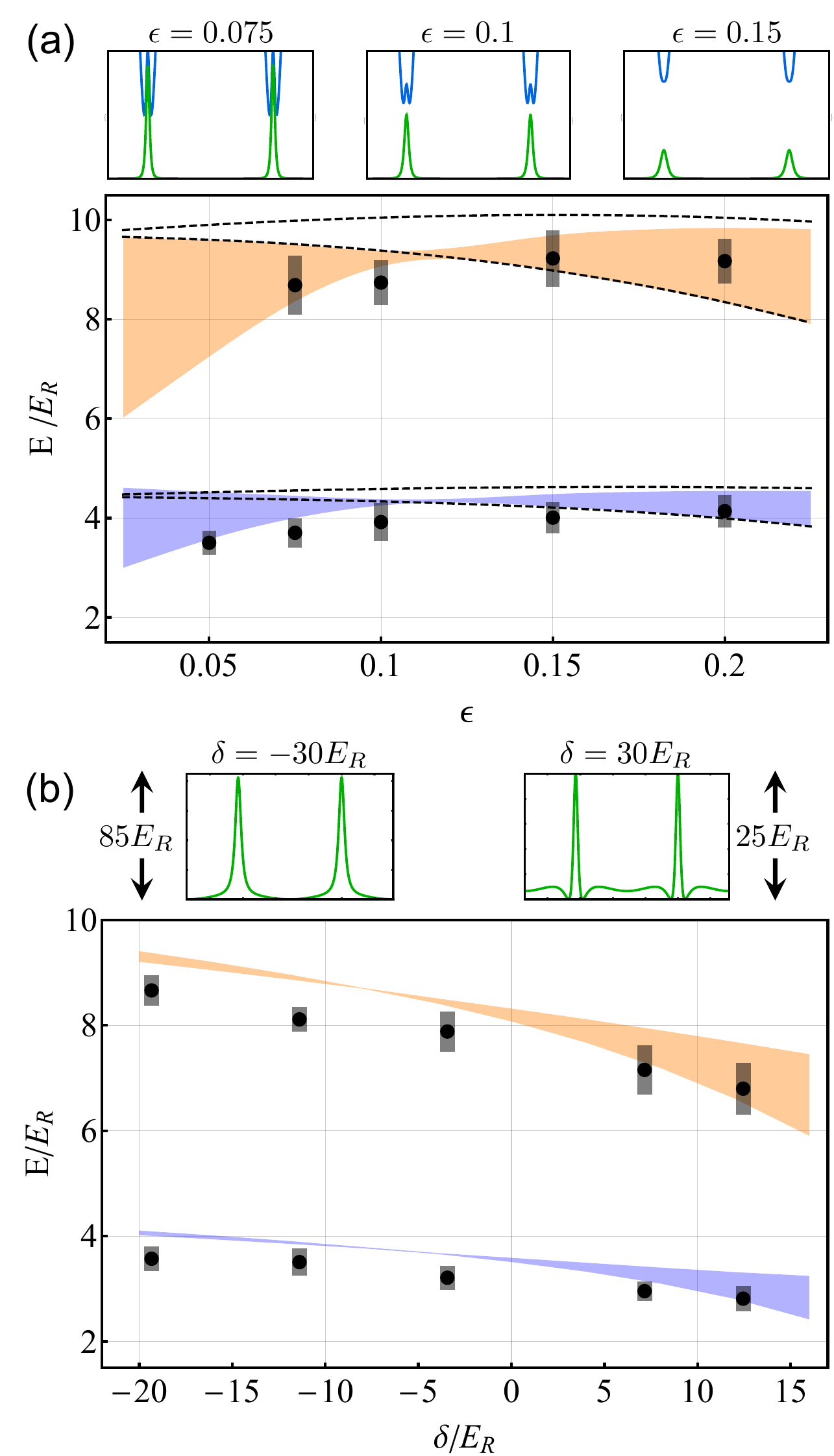}
\caption{(Color online) Band structure scalings.
Energies of the $p$- and the $d$-bands with respect to the $s$-band are plotted. 
\textbf{(a)} Vary $\epsilon$: $\Omega_c=100\Gamma$, $\Omega_p=5 \Gamma-20\Gamma$, $\Delta=22\Gamma$, and $\delta=0$. 
Dashed lines indicate the allowed transition energies predicted from modeling $V(x)$ alone, while the shaded regions are from a model including couplings to the bright states. Upper panels show representative potentials for the dark state (green) and bright state (blue). At $\epsilon= 0.075$, the bright/dark states are no longer good basis states because of the strong coupling between them.
\textbf{(b)} Vary $\delta$: $\Omega_c=70\Gamma$, $\Omega_p=10\Gamma$, $\Delta=22\Gamma$. Upper panels show calculated dark state potentials for positive and negative $\delta$. 
 }
\label{fig:band}
\end{figure}
%-------------------------------------------FIGURE 3-------------------------------------------------------------------

Another property of a KP lattice is that in the deep lattice limit, its band structure is almost independent of the barrier strength (defined as the area under the potential for a single barrier), which scales with $1/\epsilon$. The band spacings for different $\epsilon$ are plotted in Fig.~\ref{fig:band}~(a) for fixed $\Omega_c=100\Gamma$ and $\Omega_p$ varied from $5 -20\Gamma$. As expected, the band spacings are almost independent of $\epsilon$, even though the probe power varies by an order of magnitude. The upper panels of Fig.~\ref{fig:band}~(a) show the potentials of the upper bright state (blue) and dark state (green) for three $\epsilon$. For $\epsilon \leq 0.1$, mixing between $E_0(x)$ and $E_{\pm}(x)$ states introduces an avoided crossing and modifies the band structure, reducing the band spacing. For $\epsilon\sim0.1$ we realize a barrier width of 10 nm with minimal coupling to the bright state. The shaded regions are predictions based on a model that takes such bright state couplings into account, which are in better agreement with the measured band spacings, compared to the model that has no such couplings (dashed line). We attribute the discrepancy between theory and experiment to the residual polarization imperfections, calibration errors in the optical intensity, and limitations of band spectroscopy. We note that the theory predicts a vanishing band width near $\epsilon\simeq0.125$ and the growth of the bandwidth at even smaller $\epsilon$, due to the interference of the dark state and bright state mediated tunneling~\cite{supplementary}.

The data we present so far are taken under the dark state condition ($\delta=0$). Away from $\delta=0$, the state is no longer completely dark and it experiences an additional periodic potential with amplitude $\delta$~\cite{supplementary,Bienias2017} (Fig.~\ref{fig:band}~(b)). This additional potential perturbs the KP lattice and modifies the band structure. We verify this effect by measuring the band spacings as a function of $\delta$ (Fig.~\ref{fig:band}~(b)), and find it agrees with the prediction (shaded area), with the systematic deviation likely coming from the same factors as in Fig.~\ref{fig:band}~(a).

%-------------------------------------------FIGURE 4------------------------------------------------------------------
\begin{figure}[!b]
\includegraphics[width=8cm]{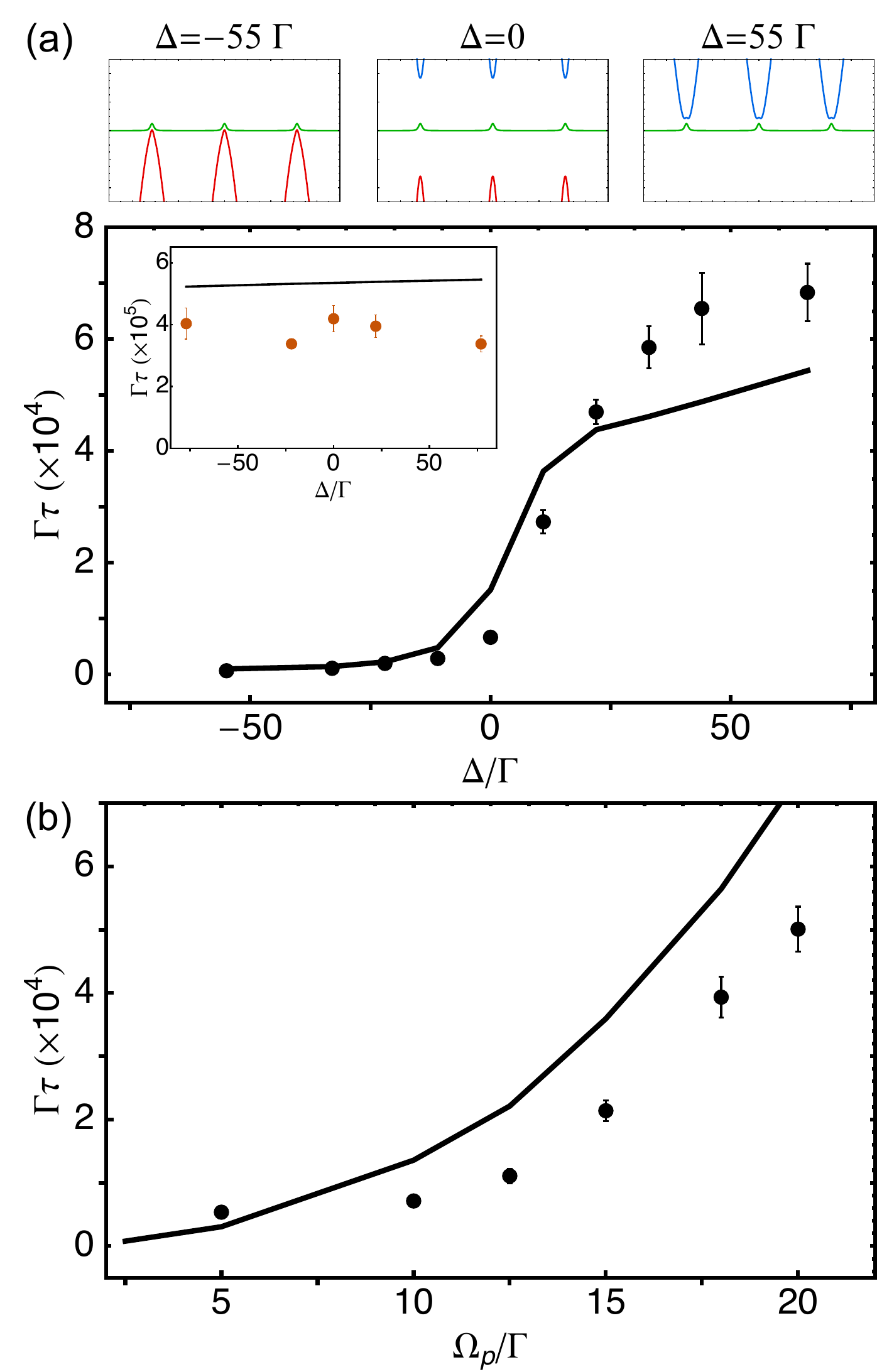}
\caption{(Color online)  
\textbf{(a)} Lifetime of dark state lattice, $\tau$, scaled by the excited state lifetime $\Gamma^{-1}$ vs. $\Delta$, with $\Omega_c=70\Gamma$, $\Omega_p=10\Gamma$, and $\delta=0$. Inset: lifetime of the dark state in spatially homogeneous control fields, with $\Omega_{c1}=35\Gamma$, $\Omega_{c2}=0$, $\Omega_p=10\Gamma$, and $\delta=0$. Upper three panels: the two bright state potentials $E_-(x)$ (red) and $E_+(x)$ (blue), and the dark state potential (green) at different $\Delta$. 
\textbf{(b)} Lifetime vs. $\Omega_p$ in a dark state lattice where $\epsilon=0.2$ and $\Delta=0$. The solid black lines are predictions scaled with a factor 2.2 (except for (a) inset, where no scaling is applied). 
The error bars represent one standard deviation uncertainty from fitting the population decay data. 
}
\label{fig:lifetime}
\end{figure}
%-------------------------------------------FIGURE 4-------------------------------------------------------------------

Finally, we study dissipation in this lattice. The non-adiabatic corrections to the BO potential that give rise to $V(x)$ also weakly couple the dark state with the bright states, which leads to light scattering, heating the atoms out of the trap. We measure the lifetime, $\tau$, in a dark state lattice (Fig.~\ref{fig:lifetime}~(a)) for different $\Delta$, and find it significantly longer for $\Delta>0$ than for $\Delta<0$. This is in contrast to an optical lattice based on ac-Stark shifts, where the heating rate is independent of the sign of $\Delta$~\cite{Gordon1980, Gerbier2010}. To intuitively understand this asymmetry, we use the model described in~\cite{Jendrzejewski2016} and note that the coupling to the bright states mostly takes place inside the barrier. An atom can scatter light by admixing with the bright states $E_{\pm}(x)$ (approximately $\Delta$ independent) or exiting into the energy-allowed $E_-(x)$ state via non-adiabatic couplings (strongly $\Delta$ dependent). The $E_-(x)$ state (red, Fig.~\ref{fig:lifetime}~(a), upper panels) contributes more to the loss, explaining the $\Delta$ asymmetry. The result of the model~\cite{supplementary} is depicted as the black solid line, with an empirical scale factor of 2.2 applied to the theory to account for the unknown relationship between the scattering rate and loss rate ($1/\tau$). The lifetime in a homogeneous control field measured when one of the $\Omega_c$ beams is blocked, is shown in Fig.~\ref{fig:lifetime}~(a) inset. The $\tau\sim4\times10^5/\Gamma$ lifetime is almost independent of $\Delta$ as theory would predict, and is 70\% of the expected lifetime due to non-adiabatic coupling to the bright states and off-resonant scattering from states outside the three-level system.

The non-adiabatic bright state coupling also leads to a counter-intuitive dependence of the dissipation on the laser power. Fig.~\ref{fig:lifetime}~(b) shows the lifetime at constant barrier height (fixed $\epsilon$) as a function of Rabi frequencies. Remarkably, the lifetime {\em{increases}} with Rabi frequency. In contrast, for a regular optical lattice at a fixed detuning the lifetime due to scattering does not improve with more laser power. For the dark state lattice, larger $\Omega_{c,p}$ increases the energy separations between BO potentials, resulting in decreased scattering. In general the lifetime improves with more laser power and at blue detuning. However, couplings to $E_+(x)$ adversely affects the barrier height (similar to the case with $\epsilon\ll1$ in Fig.~\ref{fig:band} (a)). With realistic increase in laser intensity, we can potentially improve the lifetime by an order of magnitude while maintaining the ultra-narrow barriers.

The conservative nanoscale optical potential demonstrated here adds to the toolbox of optical control of atoms, enabling experiments requiring sub-wavelength motional control of atoms. Such sharp potential barriers could be useful for the creation of narrow tunnel junctions for quantum gases~\cite{Eckel2014} or for building sharp-wall box-like traps~\cite{Gaunt2013}. In addition, spin and motional localization on small length scales can enhance the energy scale of weak, long range interactions~\cite{Lacki2016}. The dark state lattice is readily generalizable to 2D, and for example, can be used to study Anderson localization with random strength in the barrier height~\cite{Morong2015}. By stroboscopically shifting the lattice~\cite{Nascimbene2015}, the narrow barriers should enable the creation of optical lattices with spacings much smaller than the $\lambda/2$ spacing set by the diffraction limit, which would significantly increase the characteristic energy scales relevant for interaction many-body atomic systems.

\appendix
 \bibliographystyle{apsrev4-1.bst}
%\bibliography{darklatticerefs}

\section*{Acknowledgments}
We thank Victor M. Galitski for stimulating discussions and Luis A. Orozco for a close reading of the manuscript. 
Y.W., S.S., T-C. T. J.V.P., and S.L.R. acknowledge support by NSF PFC at JQI and ONR. 
P.B. and A.V.G. acknowledge support by NSF PFC at JQI, AFOSR, ARL CDQI, ARO, ARO MURI, and NSF QIS. 
M.\L. acknowledges support of the National Science Centre, Poland via project 2016/23/D/ST2/00721.
M.\L., M.A.B., and P.Z. acknowledge support from the ERC Synergy Grant UQUAM, the Austrian Science Fund through SFB FOQUS (FWF Project No. F4016-N23), and EU FET Proactive Initiative SIQS.



\section*{Supplementary material}

# Supplementary Material: Dark state optical lattice with sub-wavelength spatial structure

Y. Wang,[1] S. Subhankar,[1] P. Bienias,[1] M. Łącki,[2,3,4] T-C. Tsui,[1] M. A. Baranov,[3,4] A. V. Gorshkov,[1,5] P. Zoller,[3,4] J. V. Porto,[1] and S. L. Rolston[1]

[1]*Joint Quantum Institute, National Institute of Standards and Technology and the University of Maryland, College Park, Maryland 20740 USA*
[2]*Jagiellonian University, Institute of Physics, Łojasiewicza 11, 30-348 Kraków, Poland*
[3]*Institute for Quantum Optics and Quantum Information of the Austrian Academy of Sciences, A-6020 Innsbruck, Austria*
[4]*Institute for Theoretical Physics, University of Innsbruck, A-6020 Innsbruck, Austria*
[5]*Joint Center for Quantum Information and Computer Science,*
*National Institute of Standards and Technology and the University of Maryland, College Park, Maryland 20742 USA*
(Dated: December 1, 2017)

## I. EXPERIMENTAL TECHNIQUES

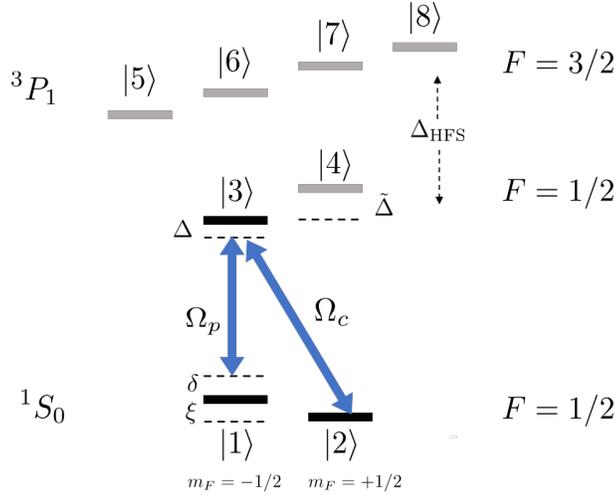

FIG. S1. Level structure of the $^1S_0$ and $^3P_1$ manifolds of $^{171}$Yb: $\delta$ is the two photon detuning; $\Delta$ is the single photon detuning; $\xi$ is the ground state Zeeman splitting and $\tilde{\Delta}$ is the Zeeman splitting in the excited state due to the external magnetic field; and $\Delta_{\text{HFS}}$ is the $^3P_1$ hyperfine splitting.

We use forced evaporation of co-trapped $^{87}$Rb to sympathetically cool $\sim 3 \times 10^5$ atoms of $^{171}$Yb to a temperature of $1.1\ T_F$. The $^{87}$Rb -$^{171}$Yb mixture is produced in a combined magnetic and multi-wavelength optical dipole trap [S1]. After reaching the final temperature, the Rb atoms are removed by ramping on the magnetic field to 12 mT. The lifetime of Yb atoms in the trap after removal of the Rb atoms is 3 s. Referring to Fig. 1c. in the main text, the final Yb trap frequencies are $\omega_x \simeq 2\pi \times 164$ Hz, $\omega_{x+y} \simeq 2\pi \times 50$ Hz, and $\omega_{x-y} \simeq 2\pi \times 155$ Hz.

The $\sigma_i^-$ and $\pi$ coupling beams at 556 nm are generated from the same laser, which is beat-note (BN) locked to a separate laser (used for laser cooling on the $^1S_0 - ^3P_1$ transition), which itself is locked to an Yb atomic saturation absorption signal. The dynamic control of the BN allows for rapid tuning of the single-photon detuning $\Delta$ of the beams. Separate acousto-optic modulators (AOM) allow for independent intensity control of all three coupling beams, as well as control over the two photon detuning $\delta$ and the phase offset between the two $\sigma_i^-$ beams ($i = 1, 2$) used for shaking the lattice. The three beams are delivered to the atoms through independent optical fibers, and have beam waists of $\sim 1$mm at the atoms.

We image the atoms using the Yb $^1S_0 - ^1P_1$ transition at 399 nm, with light generated by a frequency doubled laser system. We stabilize the seed of the imaging laser (at 800 nm) via a scanning transfer cavity lock [S2, S3] with light locked to the $5^2S_{1/2} - 5^2P_{3/2}$ transition at 780 nm in Rb as the reference. State-selectivity is achieved by imaging in a large magnetic field of 12 mT along $\hat{x}$, such that the resulting 116 MHz spacing between the Zeeman states of $^1P_1$ is sufficient to resolve the spin-dependent transitions with linewidth $\Gamma_{1P1} = 2\pi \times 30$ MHz. The imaging beam propagates along $\hat{y}$ with linear polarization along $\hat{z}$, such that it has an equal superposition of $\hat{\sigma}_+$ and $\hat{\sigma}_-$ relative to the B-field.

We measure the population in each hyperfine ground state by making the laser resonant with its respective stretched state ($^1S_0 \left| F = 1/2, m_F = \pm 1/2 \right\rangle \leftrightarrow {}^1P_1 \left| F = 3/2, m_F = \pm 3/2 \right\rangle$) during imaging.

## II. RABI FREQUENCY CALIBRATION

We calibrate the Rabi frequencies of $\sigma_i^-$ ($\Omega_{ci}$) and $\pi$ ($\Omega_p$) by measuring two-photon Raman Rabi frequencies and light-shift induced Raman detuning of the $\sigma_i^- - \pi$ pairs of beams as a function of their beam powers. When $\Delta \gg \Gamma, \Omega$, one can adiabatically eliminate the excited state, $\left| 3 \right\rangle$, and reduce the three-level system to an effective two-level system with states $\left| \vec{p} \right\rangle \left| 1 \right\rangle$ and $\left| \vec{p} + 2\hbar\delta\vec{k} \right\rangle \left| 2 \right\rangle$, resulting in the following expressions for the two-photon Rabi frequency ($\Omega_R$) and Raman detuning ($\delta_R$) [S4]

$$\Omega_R = \frac{\Omega_p \Omega_{ci}}{2\Delta} \tag{S1}$$

$$\delta_R = \delta + \frac{4\omega_R(\vec{p} \cdot \delta\vec{k} + \left| \delta\vec{k} \right|)}{\left| \delta\vec{k} \right|} + \xi - \frac{\Omega_{ci}^2}{4\Delta} + \frac{\Omega_p^2}{4\Delta} \tag{S2}$$

where $\delta\vec{k} = \vec{k}_c - \vec{k}_p$, and $\omega_R = \frac{\hbar \left| \delta\vec{k} \right|^2}{2m}$ is the Raman recoil energy. $\vec{k}_c$ and $\vec{k}_p$ are the k-vectors for one of the pump beams and probe beam respectively. For a stationary gas, the second term in (S2) averages to 0.

We extract $\Omega_R$ and $\delta_R$ by measuring the oscillation of the center of mass (COM) position of the cloud as a function of the Raman pulse time, after a 12 ms time-of-flight (TOF). In addition to transferring population between ground states, stimulated Raman transitions provide a momentum kick i.e. $\left| \vec{p} \right\rangle \left| 1 \right\rangle \leftrightarrow \left| \vec{p} + 2\hbar\delta\vec{k} \right\rangle \left| 2 \right\rangle$. This momentum kick manifests itself in TOF measurements as spatially separated momentum peaks, and the COM position of the cloud provides a measure of the ground state populations. Fig. S2(a) shows typical Rabi oscillation data, which we fit to a damped sinusoid to determine $\Omega_R$. The damping arises from the spread in the momentum distribution of the atoms.

$\Omega_R$ gives us information about the product of $\Omega_{ci}$ and $\Omega_p$, but not their absolute magnitudes. For fixed $\Delta$, $\Omega_R$ and $\delta_R = 0$, Eqs. (S1) and (S2) can be used to determine the two Rabi frequencies $\Omega_{ci}$ and $\Omega_p$ as a function of laser power. Defining $\Omega_p^2 = A_p P_p$ and replacing $\Omega_{ci} = 2\Omega_R \Delta / \Omega_p$ in Eq. (S2), we get

$$\delta = \xi - \frac{A_p P_p}{4\Delta} + \frac{\Delta \Omega_R^2}{A_p P_p} \tag{S3}$$

where $P_p$ is the power of the probe beam and $A_p$ is the constant that connects $\Omega_p$ to $P_p$.

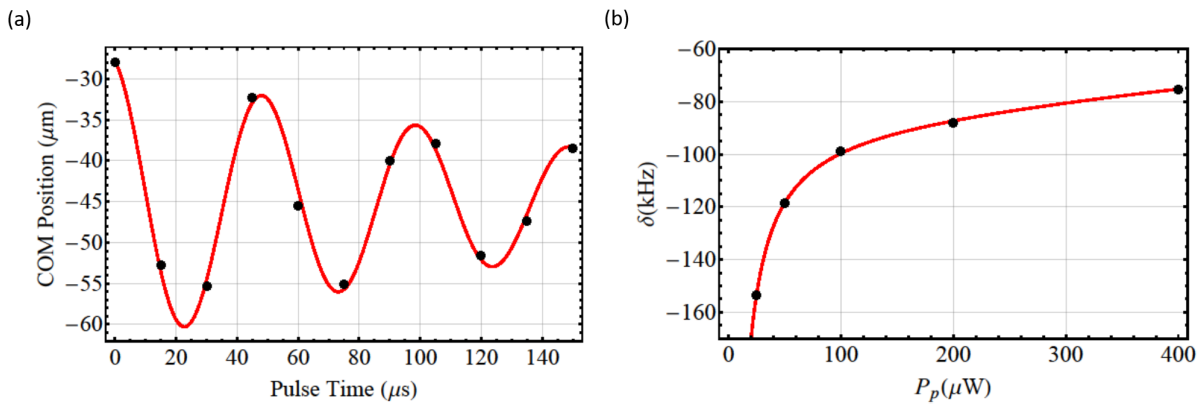

FIG. S2. (a) Example center of mass oscillation of the cloud as a function of Raman pulse time, used to calibrate Rabi frequencies. (b) Change in two-photon detuning ($\delta$) as function of the probe power, $P_p$.

Experimentally, we fix $\Delta = 40$ MHz and vary $P_{ci}$ (the power of $\sigma_i^-$ beam) and $P_p$ so as to keep $\Omega_R$ constant. We tune $\delta$ to satisfy $\delta_R = 0$, determined by maximizing the COM shift after a $\pi/2$ Raman pulse. We fit the measured

values for $\delta$ as a function $P_p$ using Eq. (S3) as the fit function with $A_p$ and $\xi$ as fit parameters. Fig. S2(b) shows the fit to data. Defining $\Omega_{ci}^2 = A_{ci}P_{ci}$ just as for the probe beam, we extract $A_{ci}$ using Eq. (S1). Given the uncertainty in the measurement of $\Omega_R$ for the two cases, we can determine the balanced condition $\Omega_{c1} = \Omega_{c2}$ to within 2%.

## III. IMBALANCED LATTICE BEAMS

To determine the potential $V(x)$ for imbalanced control beams, $\Omega_{c1} \neq \Omega_{c2}$, we consider the spatially dependent control Rabi frequency of the form

$$\Omega_c(x) = \frac{\Omega_{c1}e^{ikx} - \Omega_{c2}e^{-ikx}}{i}$$ (S4)

When $\Omega_{c1} = \Omega_{c2} = \Omega_c$, we recover $\Omega_c(x) = 2\Omega_c \sin(kx)$. The resulting Hamiltonian in the bare state basis has the form

$$H = \frac{\hbar\Omega_p}{2}\begin{pmatrix} 0 & 0 & 1 \\ 0 & 0 & s\sin(kx) + i\eta\cos(kx) \\ 1 & s\sin(kx) - i\eta\cos(kx) & 0 \end{pmatrix}$$ (S5)

where $s = \frac{1}{\epsilon} = \frac{\Omega_{c1}+\Omega_{c2}}{\Omega_p}$ and $\eta = \frac{\Omega_{c1}-\Omega_{c2}}{\Omega_p}$, and $\Delta = \Gamma = \delta = 0$. Diagonalizing $H$ gives the following normalized eigenvectors:

$$|E_0(x)\rangle = -\frac{i\eta\cos(kx) + s\sin(kx)}{\sqrt{(1+\eta^2\cos^2(kx)+s^2\sin^2(kx))}}\,|1\rangle + \frac{1}{\sqrt{(1+\eta^2\cos^2(kx)+s^2\sin^2(kx))}}\,|2\rangle$$

$$|E_-(x)\rangle = -\frac{1}{\sqrt{2(1+\eta^2\cos^2(kx)+s^2\sin^2(kx)}}\,|1\rangle + \frac{(i\eta\cos(kx)-s\sin(kx))}{\sqrt{2(1+\eta^2\cos^2(kx)+s^2\sin^2(kx))}}\,|2\rangle + \frac{1}{\sqrt{2}}\,|3\rangle$$

$$|E_+(x)\rangle = \frac{1}{\sqrt{2(1+\eta^2\cos^2(kx)+s^2\sin^2(kx)}}\,|1\rangle + \frac{(-i\eta\cos(kx)+s\sin(kx))}{\sqrt{2(1+\eta^2\cos^2(kx)+s^2\sin^2(kx))}}\,|2\rangle + \frac{1}{\sqrt{2}}\,|3\rangle$$

The resulting scalar potential for the dark state $|E_0(x)\rangle$ [S5] is

$$V(x) = \frac{\hbar^2}{2M}\sum_{j=\pm}|\langle E_j(x)|\nabla|E_0(x)\rangle|^2 = E_R\frac{s^2\cos^2(kx)+\eta^2\sin^2(kx)}{(1+\eta^2\cos^2(kx)+s^2\sin^2(kx))^2}$$ (S6)

When there is no imbalance, $\eta = 0$, we recover Eq. (1) from the main text,

$$V(x;\eta=0) = E_R\frac{s^2\cos^2(kx)}{(1+s^2\sin^2(kx))^2}$$ (S7)

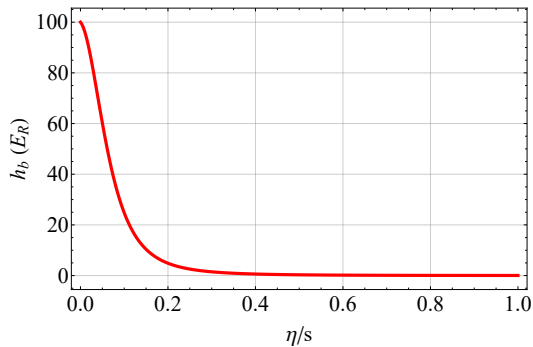

FIG. S3. Growth of the barrier with decrease in $\eta/s$.

The barrier height is given by $h_b = V(x = 0) = E_R s^2/(1 + \eta^2)^2$. In Fig. S3, we show how $h_b$ grows as the imbalance $\eta/s$ is reduced from 1 to 0 (adiabatically loading into a $100E_R$ lattice). When one of the control beams is turned off, $s^2 = \eta^2$, the potential $V(x)$ is small and spatially homogeneous. Given that our Rabi frequencies ($\Omega_{c1}, \Omega_{c2}$) are balanced to within 2%, the uncertainty in $\eta$ has negligible effect on the barrier height as $h_b \simeq E_R s^2(1 - 2\eta^2)$ for $\eta \ll 1$. We also note that the FWHM of the sub-wavelength barriers in the dark state lattice is well approximated by $0.2\lambda\epsilon$ below $\epsilon = 0.3$.

## IV. LOADING ATOMS INTO DARK STATE LATTICE AND BANDMAPPING

To adiabatically load atoms into the ground band of the lattice, we first prepare the atoms in $|1\rangle$ by optically pumping using the $\sigma_1^-$ beam with $\Delta = 12$ MHz. We then change $\Delta$ to the final desired value using the BN lock. We ramp the power of the $\sigma_1^-$ beam in 0.1 ms and hold for 0.1 ms, followed by ramping on the $\pi$ beam in 0.3 ms and holding for 0.1 ms. This adiabatically transfers the atoms in $|1\rangle$ into the dark state of the $\sigma_1^- - \pi$ beam configuration. Finally, we ramp on the $\sigma_2^-$ beam in 1 ms. As discussed in the previous section, this final step ramps the lattice imbalance from $\eta = 1$ to $\eta = 0$, smoothly transforming the potential into the lattice potential.

To find the quasimomentum distribution of the atoms in the dark lattice we perform bandmapping at the end of the experiment by adiabatically ramping down $\sigma_2^-$ in 0.5 ms (ramping down the barrier height), and then snapping off the $\sigma_1^-$ beam, the $\pi$ beam and the optical dipole trap. This maps atoms with given quasimomentum in the lattice to real momentum in free space. We then take an absorption image of this momentum distribution after 12 ms TOF.

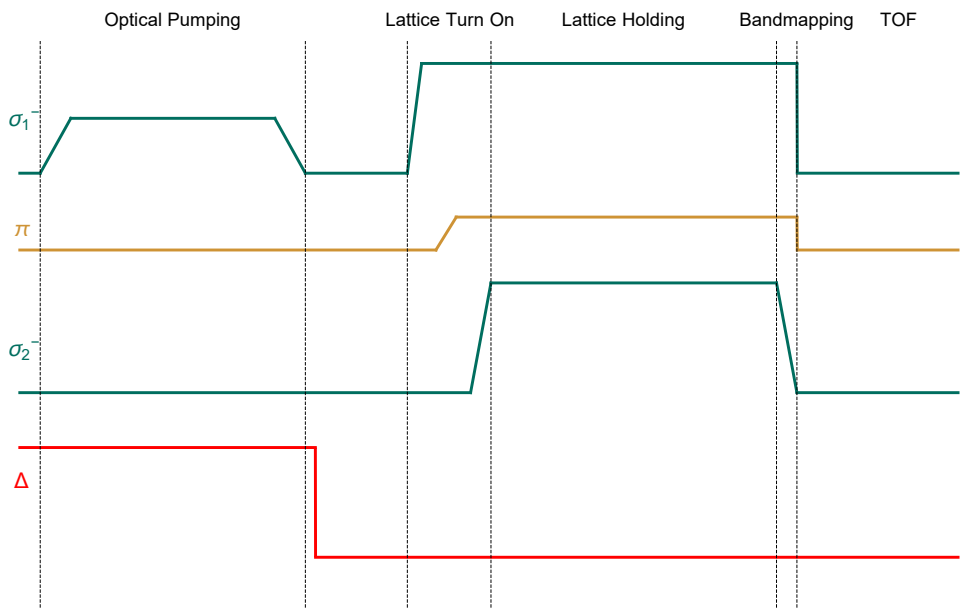

FIG. S4. Procedure to adiabatically load atoms into the ground band of the dark state lattice and performing bandmapping at the end of the sequence.

## V. POLARIZATION OPTIMIZATION

The polarization purity of the beams is sensitive to the alignment of the beams, $\vec{k}_p$ and $\vec{k}_{ci}$, with the magnetic field $\vec{B}$. When $\vec{k}_{ci}$ of the circularly polarized control beams is well aligned with $\vec{B}$, the atoms are optically pumped into the trivial dark state, $|1\rangle$. But when the magnetic field is misaligned by an angle $\theta$ from $\vec{k}_{ci}$, the resulting $\pi$ component intensity scales as $\sin^2(\theta)$. The effect of the $\pi$ component can be investigated in the context of off-resonant EIT. Due to the non-zero ground state Zeeman splitting (89 kHz), the $\pi$ component and the $\sigma^-$ component of the control beam are not on two-photon resonance which leads to photon scattering. This causes loss of atoms from the trap. By measuring atom loss as function of the strength of the transverse magnetic field component($\vec{B}_\perp$), we adjust $\vec{B}$ such that $\vec{B} \parallel \vec{k}_{c1}$. We then align $\vec{k}_{c2}$ to $\vec{k}_{c1}$ by coupling the $\sigma_2^-$ beam into the optical fiber that delivers the $\sigma_1^-$ beam to the atoms with over 50% coupling efficiency.

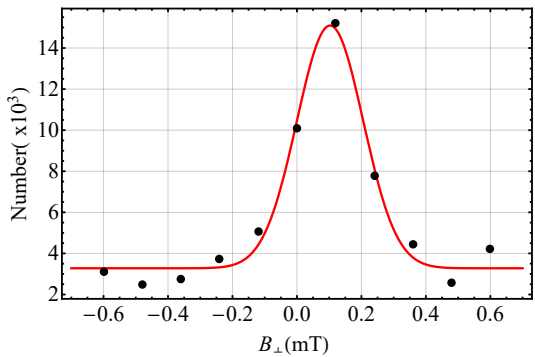

FIG. S5. Atom loss measurement to optimize polarization of $\sigma_1^-$ beam by scanning the transverse component of magnetic field, $\vec{B}_\perp$.

The polarization of the $\pi$ beam is optimized by aligning its polarization $\vec{\epsilon}_\pi$ along $\vec{B}$, $\vec{\epsilon}_\pi \times \vec{B} = 0$. This also implies $\vec{k}_p \cdot \vec{B} = 0$ where $\vec{B}$ is the magnetic field that also satisfies $\vec{B} \parallel \vec{k}_{c1}$. We use a Glan-Taylor polarizer to define $\vec{\epsilon}_\pi$. The optimal setting for the polarizer is determined by minimizing loss of atoms from the trap. To satisfy $\vec{k}_p \cdot \vec{B} = 0$, we adjust $\vec{k}_p$ such that the same $\vec{B}_\perp$ satisfies both $\vec{k}_p \cdot \vec{B} = 0$ and $\vec{B} \parallel \vec{k}_{c1}$. We do this via a scheme where we first adjust $\vec{k}_p$ and then perform the measurement we did to make $\vec{B} \parallel \vec{k}_{c1}$ (Fig.S5), but now with the $\pi$ beam. We iterate this scheme until the $\vec{B}_\perp$ we measure coincides with the $\vec{B}_\perp$ needed to satisfy $\vec{B} \parallel \vec{k}_{c1}$. When the polarization of the $\pi$ beam is not well aligned with $\vec{B}$, undesired $\sigma$ component from the $\pi$ beam causes increased atom loss from the trap. Just as in the case for $\sigma_1^-$, this atom loss can be understood in the context of off-resonant EIT. With these techniques, we accurately make $\vec{B} \parallel \vec{k}_{c1}$, $\vec{k}_p \cdot \vec{B} = 0$ and $\vec{\epsilon}_\pi \times \vec{B} = 0$ to within $\sim 5$ mrad.

## VI. LATTICE MODULATION SPECTROSCOPY

We probe the bandstructure of the dark lattice by modulating the phase of the $\sigma_2^-$ beam with respect to the $\sigma_1^-$ beam using an analog RF phase-shifter on the RF drive to the AOM for the $\sigma_2^-$ beam. We transfer atoms from the ground band to the first excited band by shaking the lattice at the appropriate frequency for 200 $\mu$s and an amplitude of $\lambda/8$. We then perform bandmapping by ramping down one of the lattice beams in 0.5 ms. We measure the population in each band as a function of modulation frequency and fit a Gaussian to the excited population to extract the transition frequencies.

Since the short lattice shaking pulse is Fourier broadened, we are unable to resolve the quasimomentum dependence of the band excitations when we scan the frequency. We also notice that in order to significantly populate excited bands, a lattice shaking pulse with large amplitude is needed, in contrast to standard sinusoidal lattices where the curvature of band dispersion can be easily resolved with small shaking amplitudes and longer pulse times. In standard sinusoidal optical lattices, the Rabi frequency between a pair of bands is proportional to the shaking amplitude and the lattice depth, and a small shaking amplitude in deep lattices can significantly populate the excited bands. This is in contrast with the KP lattice, where the Rabi frequency is proportional to shaking amplitude, but independent of the barrier height. Larger shaking amplitudes are required to transfer substantial population to the excited bands. In addition, since the wavefunction in the dark state lattice goes to zero at the barriers and the potential is flat elsewhere (where most of the wavefunction lives), the barrier excursion during a lattice shaking pulse needs to be large to diabatically distort the wavefunction between the barriers and hence cause band excitations.

In Figs. 3(a) and 3(b) of the main text, we note that the frequencies of the transitions ($s \rightarrow p$ and $s \rightarrow p \rightarrow d$) measured are slightly lower than the shaded regions. The dipole nature of the coupling due to lattice shaking allows only transitions between bands of opposite parity i.e. $s \leftrightarrow p$ and $p \leftrightarrow d$. The population in the $d$ band arises after some population has been transferred into the $p$ band from the $s$ band, which may affect the resonance spectrum. To better understand this potential systematic redshift of the measured transition frequencies, we employ a model where we map the first three bands ($s, p, d$) of the dark-lattice to an ensemble of discrete three-level ladder-type systems where the energy spacings trace the curvatures of the bands sampled at various values of $q$. We solve the full time-dependent Hamiltonian with an appropriate scaling factor for the drive strength and fit the ensemble average of the population in the excited states as a function of lattice shaking frequency to estimate the transition frequencies. These estimated transition frequencies are systematically lower by $\sim E_R/3$ than the average transition frequency between the bands

which, has the same sign as the redshifts observed but is too small to account for all of the shift.

## VII. MULTI-LEVEL MODEL

Fig. S1 depicts the three hyperfine states that constitute the $\Lambda$-system consisting of $|1\rangle$, $|2\rangle$, and $|3\rangle$ (black) along with five hyperfine states (grey) that the $\Lambda$ system is coupled to off-resonantly. Couplings to these off-resonant states affect the bandstructure of the $\Lambda$-system. We account for the effect of these off-resonant couplings by adiabatically eliminating the five far off-resonant hyperfine states, which to second ordeer shifts the energies of the ground states of the $\Lambda$-system, $|1\rangle$ and $|2\rangle$, by $\delta_1(x)$ and $\delta_2(x)$ respectively.

The Hamiltonian for our effective $\Lambda$-system is as follows.

$$H = -\frac{\hbar^2 \partial_x^2}{2m} + \hbar \begin{pmatrix} \delta_1(x) & 0 & \frac{\Omega_p}{2} \\ 0 & \delta_2(x) & \frac{\Omega_c(x)}{2} \\ \frac{\Omega_p}{2} & \frac{\Omega_c(x)}{2} & -(\Delta + i\frac{\Gamma}{2}) \end{pmatrix} \tag{S8}$$

Here we use the same treatment to solve this non-hermitian Hamiltonian, as explained in [S6]. As $H$ has the periodicity of $\lambda$, we use Bloch's theorem to define the wavefunction that solves $H$ as $\vec{\Phi}(x) = (u_1(x), u_2(x), u_3(x))^T$, where $u_j(x) = \sum_{n=-N}^{n=N} C_{j,q,n} e^{i(q + \frac{2n\pi}{\lambda})x}$ for $j = 1, 2, 3$. Here $u_1(x), u_2(x)$, and $u_3(x)$ represent the atomic wavefunctions for states $|1\rangle$, $|2\rangle$ and $|3\rangle$ respectively, and $q$ is the quasimomentum such that $q \in [-\pi/\lambda, \pi/\lambda]$. The basis truncation at $N$ must be chosen large enough to correctly capture the non-adiabatic couplings between bright and dark Born-Oppenheimer states at high momenta, and for the typical values of the parameters used in the experiment (e.g. $\Omega_p = 5 - 30\Gamma$, $\Omega_c = 70 - 100\Gamma$, and $\Delta = 2\pi \times 4$MHz), we find $N = 105$ to be sufficient to simulate the bandstructure.

The expressions for the ac-Stark shifts, $\delta_1(x)$ and $\delta_2(x)$, due to off-resonant Rabi couplings are as follows.

$$\delta_1(x) = \delta - \frac{\Omega_p^2}{2\Delta_{\text{HFS}}} - \frac{3\Omega_c(x)^2}{8\Delta_{\text{HFS}}} \tag{S9}$$

$$\delta_2(x) = -\frac{\Omega_p^2}{4\Delta} - \frac{\Omega_p^2}{2\Delta_{\text{HFS}}} - \frac{\Omega_c(x)^2}{8\Delta_{\text{HFS}}} \tag{S10}$$

Substituting the expressions for $\delta_1(x)$ and $\delta_2(x)$ in Eq. (S8) and using the Bloch ansatz we calculate the bandstructure for the $\Lambda$ system.

The effects of the off-resonant couplings on the dark state ($|E_0(x)\rangle$) can also be understood in the context of perturbation theory. With the perturbing Hamiltonian given by

$$H_{\text{pert.}} = \hbar \begin{pmatrix} \delta_1(x) & 0 & 0 \\ 0 & \delta_2(x) & 0 \\ 0 & 0 & 0 \end{pmatrix} \tag{S11}$$

the first order correction to $V(x)$ is:

$$V_{\text{pert.}}(x) = \langle E_0(x)| H_{\text{pert.}} |E_0(x)\rangle = \hbar \left( \frac{\delta_2 \Omega_p^2}{\Omega_p^2 + \Omega_c^2(x)} + \frac{\delta_1 \Omega_c^2(x)}{\Omega_c^2(x) + \Omega_p^2} \right) \tag{S12}$$

where $|E_0(x)\rangle = \frac{-\Omega_c(x)|1\rangle + \Omega_p|2\rangle}{\sqrt{\Omega_c^2(x) + \Omega_p^2}}$. With independent control over the two-photon detuning ($\delta$), we are able to engineer sub-wavelength traps as depicted in the upper panel of Fig. 3(b) in the main text.

We also note an interesting feature upon simulating the bandstructure for various values of $\epsilon$. As $\epsilon$ is decreased from 0.2 to 0.05, we observe that the curvature of the band inverts as shown in Fig. S6. For reference, we also present the bandstructure calculated for just the potential $H = \frac{p^2}{2m} + V(x) + V_{\text{pert.}}(x)$ for each $\epsilon$, represented as black dashed lines. At a particular value of $\epsilon$, which in this case is $\simeq 0.125$, the bandwidth vanishes due to destructive interference between the normal hopping within dark state and the upper bright state assisted hopping implying no tunneling despite barriers having finite width and height.

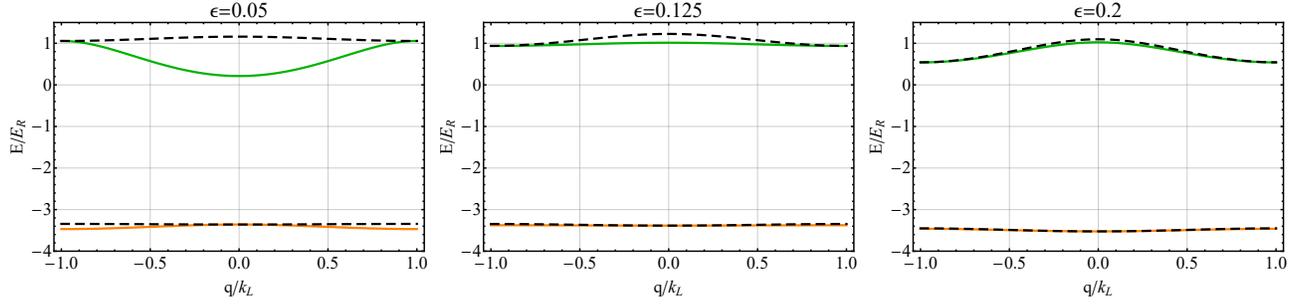

FIG. S6. Change in curvatures of the first two bands of the dark state lattice for different $\epsilon$. Colored lines represent bands for the $H$ given by Eq. (S8) and black dashed lines represent bands for $H = \frac{p^2}{2m} + V(x) + V_{\text{pert.}}(x)$.

## VIII. ESTIMATE OF LOSSES BASED ON TRANSMISSION THROUGH A SINGLE BARRIER

In this section, we provide estimates of losses present in our system, based on the analysis of scattering from a single barrier created by $\Omega_c = \Omega_p x/w$, where $w = \epsilon\lambda/(2\pi)$ [S7]. The advantage of this approach is that we can analytically calculate $P_{\text{NA}}$, the probability to lose an atom into state $|E_-\rangle$ from state $|E_0\rangle$ due to non-adiabatic coupling between the states, as well as provide intuitive explanations for $P_{\text{sc}}$, the probability of scattering a photon from state $|3\rangle$. This approach is discussed in Ref. [S7], where $P_{\text{NA}}$ and $P_{\text{sc}}$ are calculated for $\Delta = 0$. An alternative complementary approach for the estimation of losses is discussed in section IX.

First, we estimate $P_{\text{sc}}$ for the case when the detuning $\Delta = 0$. We follow the physical argument from Ref. [S7] to estimate the probability $P_e$ of being in the excited state $|3\rangle$: the dark state probability at $x = 0$ is $\sim |t|^2 \sim E/E_w$, where $E_w = \hbar^2/(2mw^2)$ is the barrier height as well as the strength of the non-adiabatic coupling. The admixture of $|E_-\rangle$ and $|E_+\rangle$ (and hence of $|3\rangle$) is simply $\sim (E_w/(\Omega_p/2))^2$, where $E_w$ plays the role of an effective Rabi frequency and $\Omega_p/2$ plays the role of detuning. Multiplying the two together we obtain $P_e \sim \frac{4EE_w}{\Omega_p^2}$. Then the probability of scattering a photon in a single pass is $P_{\text{sc}} \sim \Gamma P_e 2w/v$, where $2w/v$ is the approximate crossing time for atoms having velocity $v = \sqrt{2E/m}$. In order to estimate the impact of these losses on dark-state lifetime in the lattice, we multiply $P_{\text{sc}}$ by the rate $R_{\lambda/2} = v/(\lambda/2)$ with which atoms scatter from the barriers separated by $\lambda/2$. This leads to photon scattering rate $\gamma_{\text{sc}} = c_{\text{sc}} P_{\text{sc}} R_{\lambda/2}$, where the prefactor $c_{\text{sc}}$ is on the order of unity. We have confirmed the validity of this expression by numerically calculating the photon scattering rate $\gamma_{\text{sc}}$ from a single barrier as

$$\gamma_{\text{sc}} = \Gamma \frac{\int_{-2w}^{2w} |\psi_e(x)|^2}{\int_0^{\lambda/2} |\psi_{E_0}(x)|^2}, \tag{S13}$$

where $\psi_e(x)$ and $\psi_{E_0}(x)$ are the amplitudes to be in states $|3\rangle$ and $|E_0\rangle$, respectively, for and incoming dark-state plan wave. From comparison with numerical results, we find that the prefactor $c_{\text{sc}}$ is approximately 0.9. The second imperfection comes from the nonzero probability $P_{\text{NA}}$ of losing the atom into the open $|E_-\rangle$ channel due to the nonadiabatic coupling between $|E_0\rangle$ and $|E_-\rangle$. Based on Ref. [S7], we know that $P_{\text{NA}} \approx 1.37\sqrt{E/E_w}e^{-1.75\sqrt{\Omega_p/(2E_w)}}$. In order to estimate the impact of these losses on dark-state lifetime in the lattice, we multiply $P_{\text{NA}}$ by the rate $R_{\lambda/2}$, which leads to the rate $\gamma_{\text{NA}} = 2P_{\text{NA}}\sqrt{2E/m}/\lambda$ of losing atoms into state $|E_-\rangle$.

Using the expressions for $\gamma_{\text{sc}}$ and $\gamma_{\text{NA}}$ found above, we can estimate the decay rate $\gamma_n$ of the $n$-th Bloch band dark state by evaluating $\gamma_{\text{sc}} + \gamma_{\text{NA}}$ at $E = E_n$. In order to test the relation between our barrier-based results and the results from a direct lattice calculation more quantitatively, we consider $\gamma_1 = \gamma_{1,\text{sc}} + \gamma_{1,\text{NA}}$ for the lowest Bloch band (subscript 1 indicates the number of the band) for the parameters as in Fig. 3(left) from the supplement of Ref. [S6]. We find good agreement between the lattice-based calculation and the barrier-based calculation, see Fig. S7. Moreover, we see that $\gamma_{1,\text{NA}}$ is much smaller than $\gamma_{1,\text{sc}}$ for the parameters considered; therefore the scaling with $\Omega_p$ is given by the scaling of $\gamma_{1,\text{sc}}$. Since, for fixed $n$, all terms except $P_e$ are $\Omega_p$-independent, the scaling of $\gamma_1$ with $\Omega_p$ is governed by the scaling of $P_e$ with $\Omega_p$. This gives us a physical explanation for the numerically observed scaling in Fig. 3(left) in the supplement of Ref. [S6], as well as for the scaling of our experimental results, which we comment on below.

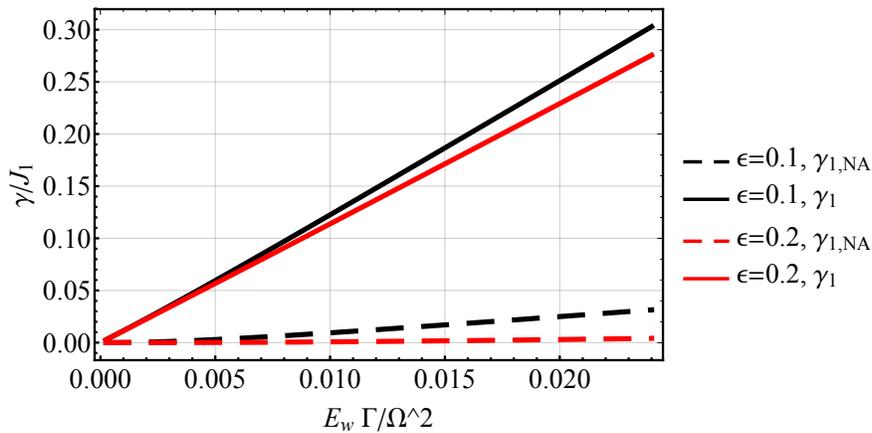

FIG. S7. We plot analytical results for the total decay rate $\gamma_1 = \gamma_{1,\mathrm{NA}} + \gamma_{1,\mathrm{sc}}$ and the non-adiabatic decay rate $\gamma_{1,\mathrm{NA}}$ for the first Bloch band in units of $J_1 = 2\epsilon E_R/\pi^2$ as a function of $E_w\Gamma/\Omega_p^2$ for $\Delta = 0$, $\Gamma = 600E_R$, and two different values of $\epsilon$. Note that the prefactor $c_{sc}$ is fixed using the numerics for a single barrier. For comparison with analogous results for a direct lattice calculation, see Fig. 3(left) in the Supplementary materials for Ref. [S6]. The numerics for a barrier gives the decay rate $\gamma_1$ approximately two times greater than the decay from lattice calculations. Note that, for $\epsilon = 0.1$, losses due to $P_{\mathrm{NA}}$ become non-negligible.

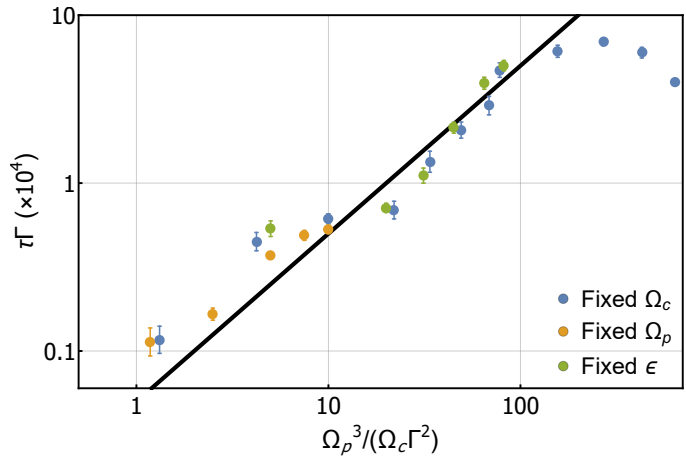

FIG. S8. Lifetime plotted on a log-log scale against the parameter $\Omega_p^3/(\Omega_c\Gamma^2)$ under three conditions: keep $\Omega_c = 100\Gamma$ and scan $\Omega_p$ from 5 to 40 $\Gamma$ (blue); keep $\Omega_p = 5\Gamma$ and scan $\Omega_c$ from 12.5 to 100 $\Gamma$ (orange); and keep $\epsilon = 0.2$ fixed and scan $\Omega_p$ from 2.5 to 20 $\Gamma$ (green). A linear fit $\tau\Gamma \propto \Omega_p^3/(\Omega_c\Gamma^2)$ is also shown (black line).

In Fig. S8, we plot the experimentally observed lifetime as a function of the parameter $\Omega_p^3/(\Omega_c\Gamma^2)$ for different cases where we vary $\epsilon$, $\Omega_p$, and $\Omega_c$. We checked based on barrier numerics for the complete data set that $\tau$ has a greater or equal contribution from $\gamma_{sc}$ than from $\gamma_{\mathrm{NA}}$. This explains the observed linear scaling over orders of magnitudes of $\tau\Gamma$ with $\Omega_p^3/(\Omega_c\Gamma^2)$. This linear scaling is plotted as a black line in Fig. S8.

Finally, we comment on the dependence of $\gamma_{sc} + \gamma_{\mathrm{NA}}$ on $\Delta$. In this case, both $P_{sc}$ and $P_{\mathrm{NA}}$ are relevant, and the interplay between them describes the results shown in Fig. 4(a) in the main text. For $\Delta < 0$, $P_{\mathrm{NA}}$ gives a leading contribution to $\tau$, whereas, for $\Delta > 0$, $P_{\mathrm{NA}}$ is negligible and the leading contribution comes from $P_{sc}$. $P_{\mathrm{NA}}$ strongly depends on $\Delta$, whereas $P_{sc}$ is only very weakly dependent on $\Delta$. This explains the observed asymmetry in the dependence of losses on $\Delta$, as explained in the main text. Finally, we numerically confirmed that the results (not shown) based on barrier calculations describe well the quantitative behavior of the losses as a function of $\Delta$ and quantitatively agree with the full lattice calculations. The full lattice analysis is presented in the next section.

## IX. THEORY FOR LIFETIME MEASUREMENTS BASED ON PERTURBATIONS TO THE DARK STATE

Here we use a direct lattice-based calculation to explain the lifetime in the context of perturbations to the dark state eigenfunction induced by couplings to the bright channels and off-resonant couplings to states outside the $\Lambda$-system (Fig. S1). The imaginary part of the dark state eigenenergy determines the scattering rate of the atoms which determine their lifetime in the trap up to a scaling factor. The couplings between the dark and bright channels are small when the energy separations between them are large. As we change $\Delta$, $\Omega_p$, and $\Omega_c$, we alter the admixture of $|3\rangle$ into the dark state and hence the amount of photon scattering. First we discuss the measured lifetime in the context of two beam EIT (inset of Fig. 4(a) in the main text) and then explain the lifetime measurements in the dark state lattice when we scan $\Delta$ and Rabi frequencies $\Omega_p$ and $\Omega_c$ (Fig. 4(a) and Fig. 4(b) in the main text).

When the off-resonant couplings to the $\Lambda$-system are neglected the Hamiltonian is given by:

$$H = -\frac{\hbar^2}{2m}\left(\frac{\partial^2}{\partial x^2} + \frac{\partial^2}{\partial y^2} + \frac{\partial^2}{\partial z^2}\right) + \hbar \begin{pmatrix} 0 & 0 & \Omega_p e^{iky}/2 \\ 0 & 0 & \Omega_c(x)/2 \\ \Omega_p e^{-iky}/2 & \Omega_c^*(x)/2 & -\Delta' \end{pmatrix}$$ (S14)

with $\Delta' = \Delta + \frac{i\Gamma}{2}$. In the experiment $\Omega_c(x)$ can be either $\Omega_c(x) = \Omega_c e^{ikx}$ with lasers in the homogeneous EIT configuration, or $\Omega_c(x) = \Omega_c \sin kx$ for standing wave case.

### A. Homogenous EIT configuration

We first consider the two-beam EIT configuration. In this configuration the non-hermitian Hamiltonian (S14) can be exactly diagonalized by finding a proper biorthogonal basis. The right eigenvectors are indexed by the 3D momentum $\vec{q}$, and are of the form:

$$e^{i\vec{q}\cdot\vec{r}}(ae^{iky}|1\rangle + be^{ikx}|2\rangle + c|3\rangle)$$ (S15)

where $(a, b, c)^\dagger$ is a right eigenvector of the matrix:

$$H_{\vec{q},k} = \begin{pmatrix} \frac{\hbar^2}{2m}(q^2 + 2kq_y + k^2) & 0 & \hbar\Omega_p/2 \\ 0 & \frac{\hbar^2}{2m}(q^2 + 2kq_x + k^2) & \hbar\Omega_c/2 \\ \hbar\Omega_p/2 & \hbar\Omega_c/2 & -\hbar\Delta' + \frac{\hbar^2 q^2}{2m} \end{pmatrix}$$ (S16)

with $q^2 = q_x^2 + q_y^2 + q_z^2$.

The Hamiltonian $H_{\vec{q},k}$ can be diagonalized analytically, however the resulting expressions are complicated. For the experimental parameters, both $\hbar^2 q^2/2m$ and the recoil energy $\hbar^2 k^2/2m$ are much smaller than $\hbar\Omega_c$ and $\hbar\Omega_p$, defining small parameters with respect to which one can expand. Expanding up to the second order we get the following expression for the energy of the dark state:

$$E_0 \approx \frac{\hbar^2 q^2}{2m} + \frac{\hbar^2}{2m}\left(k^2 + \frac{2kq_y\Omega_c^2}{\Omega_c^2 + \Omega_p^2} + \frac{2kq_x\Omega_p^2}{\Omega_c^2 + \Omega_p^2}\right) - \left(\frac{\hbar^3}{m^2}\right)\frac{4\Delta\Omega_c^2\Omega_p^2 k^2 \left(q_y - q_x\right)^2}{\left(\Omega_c^2 + \Omega_p^2\right)^3} - \left(\frac{\hbar^3}{m^2}\right)\frac{2i\Gamma\Omega_c^2\Omega_p^2 k^2 \left(q_y - q_x\right)^2}{\left(\Omega_c^2 + \Omega_p^2\right)^3},$$ (S17)

with imaginary part Im$E_0$ $\Delta$-independent. The last two terms in (S17) appear then as a second order correction. It should be mentioned that the imaginary term in Eq. (S17) is a sum of $\Delta$-dependent contributions from the upper and from the lower bright states. In their sum, however, the $\Delta$-dependence disappears.

The other loss mechanism results from off-resonant couplings to the states beyond the $\Lambda$-system ($|4\rangle$, $|5\rangle$, $|6\rangle$, and $|7\rangle$ in Fig. S1). This additional loss rate is given by

$$\Gamma_{\text{offres.}} = \Gamma\frac{\Omega_p^4}{4\tilde{\Delta}^2\left(\Omega_c^2 + \Omega_p^2\right)} + \Gamma\frac{5\Omega_c^2\Omega_p^2 + 3\Omega_c^4 + 4\Omega_p^4}{8\Delta_{\text{HFS}}^2\left(\Omega_c^2 + \Omega_p^2\right)},$$ (S18)

and is approximately $\Delta$-independent. For the parameters used in the experiment for the EIT configuration we find $\Gamma_{\text{offres.}}$ to be approximately a factor of two smaller than the loss rate from Eq. (S17). We conclude that the combined loss rate in the homogeneous EIT configuration is approximately $\Delta$-independent, in agreement with the results shown in the inset of Fig. 4(a).

## B. Standing wave case

We now consider the standing-wave case. In contrast to the homogeneous EIT configuration, the standing wave case does not admit an exact analytic treatment. Although the Hamiltonian (S14) can be diagonalized numerically, to support our numerical results and to get an insight into underlying loss mechanisms, we present below the analysis based on the Born-Oppenheimer (BO) approach (as in Ref. [S6]).

We first diagonalize the atomic part of the non-Hermitian Hamiltonian (S14) (the last term) to find its BO eigenstates: The dark state right eigenvector

$$|E_0(y,x)\rangle = \frac{1}{N_0} \left[ (-\Omega_p) |2\rangle + \Omega_c(x) e^{iky} |1\rangle \right], \quad N_0 = \sqrt{\Omega_c^2(x) + \Omega_p^2}$$

and the two bright states

$$|E_\pm(y,x)\rangle = \frac{1}{N_\pm} \left[ \Omega_c(x)|2\rangle + \left( \Delta' \pm \sqrt{\Omega_c^2(x) + \Omega_p^2 + \Delta'^2} \right) |3\rangle + \Omega_p e^{iky}|1\rangle \right],$$

where $N_\pm = \sqrt{2}\sqrt{\Omega_c^2(x) + \Omega_p^2 + \Delta' \left( \Delta' \mp \sqrt{\Omega_c^2(x) + \Omega_p^2 + \Delta'^2} \right)}$.

The eigenfunctions of the Hamiltonian (S14) can be written as

$$\psi(x,y,z) = e^{iq_z z + iq_y y} \left( f_0(x)|E_0(y,x)\rangle + f_+(x)|E_+(y,x)\rangle + f_-(x)|E_-(y,x)\rangle \right),$$

and the Hamiltonian for the wave functions $f_0(x)$ and $f_\pm(x)$ in the BO basis takes the form (see Ref. [S6])

$$
\begin{aligned}
\mathcal{H} = &-\frac{\hbar^2 \partial_x^2}{2m} + \frac{\hbar^2 q_y^2}{2m} + \frac{\hbar^2 q_z^2}{2m} + \begin{pmatrix} 0 & 0 & 0 \\ 0 & E_+ & 0 \\ 0 & 0 & E_- \end{pmatrix}. \\
&+ \frac{\hbar^2}{2m} \frac{\Omega_c'^2(x)\Omega_p^2}{\left(\Omega_c^2(x) + \Omega_p^2\right)^2} \begin{pmatrix} 1 & 0 & 0 \\ 0 & M_+^2 + C^2 & 0 \\ 0 & 0 & M_-^2 + C^2 \end{pmatrix} \\
&- \frac{i\hbar}{2m} \left( \partial_x \hat{A} + \hat{A} \partial_x \right) \\
&- \frac{\hbar^2}{2m} \frac{\Omega_c'^2(x)\Omega_p^2}{\left(\Omega_c^2(x) + \Omega_p^2\right)^2} \begin{pmatrix} 0 & CM_- & -CM_+ \\ CM_- & 0 & -M_+M_- \\ -CM_+ & -M_+M_- & 0 \end{pmatrix} \\
&+ \frac{\hbar^2}{2m} \begin{pmatrix} D_{00} & D_{0-} & D_{0+} \\ D_{0-} & D_{--} & D_{+-} \\ D_{0+} & D_{-+} & D_{++} \end{pmatrix},
\end{aligned} \tag{S19}
$$

where

$$E_\pm = \frac{1}{2} \left( -\Delta' \pm \sqrt{\Omega_c^2(x) + \Omega_p^2 + \Delta'^2} \right) \tag{S20}$$

and the matrix $\hat{A}$ has the form

$$\hat{A} = -i\hbar \frac{\Omega_c'(x)\Omega_p}{\Omega_c^2(x) + \Omega_p^2} \begin{pmatrix} 0 & -M_+ & -M_- \\ M_+ & 0 & -C \\ M_- & C & 0 \end{pmatrix}.$$

The coefficients

$$C = \Delta' \frac{\Omega_c(x)}{2\Omega_p} \frac{\Omega_c^2(x) + \Omega_p^2}{\Omega_c^2(x) + \Omega_p^2 + \Delta'^2}$$

and

$$M_\pm = \left[1 + \frac{2E_\pm(x)}{\Omega_c^2(x) + \Omega_p^2}\right]^{-1/2}.$$

determine couplings between the dark and the bright state manifold, which takes place in the region of subwavelength peaks. They also result in the energy corrections to the bright states [second line, Eq (S19)], which are typically much smaller than $E_\pm$. The matrix in the last line of (S19) originates from the $y$-dependent phase of $\Omega_p$ and has elements

$$D_{00} = \frac{\Omega_c^2(x)k(k+2q_y)}{N_0^2},$$

$$D_{0\pm} = \frac{\Omega_c(x)\Omega_p k(k+2q_y)}{N_0 N_\pm},$$

$$D_{\pm\pm} = \frac{\Omega_p^2 k(k+2q_y)}{N_\pm^2},$$

$$D_{\mp\pm} = \frac{\Omega_p^2 k(k+2q_y)}{N_\pm N_\mp}.$$

Note that $D_{00}$ gives a correction to the dark state non-adiabatic potential, Eq. (1) in the main text, which is small under the conditions of the experiment.

Considering the first two terms in (S19) as a zero-order Hamiltonian (where the dark- and bright state BO channels are decoupled), we define the dark state eigenfunctions $\psi_{(\vec{q},n)}^{E_0}(x,y,z)$ and bright states ones $\psi_{(\vec{q},n)}^{E_\pm}(x,y,z)$, where $q_z$ and $q_y$ are the transverse momenta, $q_x \in [-\pi/\lambda, \pi/\lambda]$ is the lattice quasimomentum, and $n$ being the band index. The corresponding energies are $E_{0,(\vec{q},n)}$ and (complex) $E_{\pm,(\vec{q},n)}$, respectively. Note that $q_x$, $q_y$, and $q_z$ are conserved quantum numbers even when the couplings between the BO channels are taken into account, in contrary to $n$.

We can now consider the couplings between the BO channels [other terms in the Hamiltonian (S19)] perturbatively assuming the gaps between the BO states being much larger than the coupling matrix elements. The finite lifetime of the dark state $\psi_{(\vec{q},n)}^{E_0}$ is given by the imaginary part of the corresponding eigenenergy, which appears in the second order contribution

$$\delta^{(2)}E_{0,(\vec{q},n)} = \sum_{\sigma=\pm}\sum_m \frac{{}_L\langle\psi_{(\vec{q},n)}^{E_0}|\mathcal{H}|\psi_{(\vec{q},m)}^{E_\sigma}\rangle_{RL}\langle\psi_{(\vec{q},m)}^{E_\sigma}|\mathcal{H}|\psi_{(\vec{q},n)}^{E_0}\rangle_R}{E_{0,(\vec{q},n)} - E_{\sigma,(\vec{q},m)}}. \tag{S21}$$

To analyze the $\Delta$-dependence of the imaginary part of $\delta^{(2)}E_{0,(\vec{q},n)}$, we first notice that the largest contributions to the matrix elements $\langle\psi_{(\vec{q},m)}^{E_\sigma}|\mathcal{H}|\psi_{(\vec{q},n)}^{E_0}\rangle$ from the spacial integrals come from the states with eigenenergies close to min Re$\epsilon_+$ for the upper bright states, and close to max Re$\epsilon_-$ for the lower ones. Matrix elements with other states are small because their spatial wave functions are either rapidly oscillating or exponentially suppressed in the coupling regions.

In the red detuned case $\Delta < 0$, the upper bright states are pushed away from the dark ones. This leads to the decrease of the coupling matrix elements with the relevant states because the amplitudes of the internal states $|1\rangle$ and $|2\rangle$ in the upper bright state decrease in favor of $|3\rangle$. On the other hand, the relevant lower bright states are pushed towards the dark ones, with increasing amplitudes of $|1\rangle$ and $|2\rangle$ in its internal structure. In the experiment we have $|\Delta| \lessapprox \Omega_c$ and, as a result, Im $E_{-,\vec{q},n}$ remains approximately unchanged and close to $-\Gamma/4$ for the states close to max Re$\epsilon_-$. Therefore, the dominant contribution to Im$\delta^{(2)}E_{0,(\vec{q},n)}$ comes from the lower bright states and increases with $|\Delta|$, resulting in the decrease of the life-time observed in the experiment, see Fig. 4(a).

In the opposite case $\Delta > 0$, the contribution of the lower dark states decreases, while the contribution of the upper bright states increases and becomes dominant. Here, however, the decrease of imaginary part of $E_{+,(\vec{q},m)}$ dominates over the growth of its real part and over weak increase of the coupling matrix elements, such that Im$\delta^{(2)}E_{0,(\vec{q},n)}$ decreases with increasing $\Delta$ in agreement with the experimental observations.

Finally, we mention that the life-time is an increasing function of Rabi frequencies $\Omega_c, \Omega_p$, see Fig. 4(b). This is due to the increase of the gaps to the bright states, while the couplings matrix elements remain practically unchanged.

Connecting Eq. (S21) to the population decay rate is done in the following way: we assume a homogeneous trapped gas with a Fermi-Dirac distribution, where the chemical potential is chosen to fit the total number of particles. For each band index $n$ and quasimomentum $\vec{q}$ the decay rate $\Gamma_{(\vec{q},n)}$ is evaluated by finding the imaginary part of the eigenenergy in Eq. (S21). Since most of the atoms occupy the lowest band of the lattice, we average $\Gamma_{(\vec{q},n)}$ over the lowest three bands to find the average decay rate, which is justified by the fact that the third band almost gives no contribution ($<5\%$). We use the experimentally measured atom number, temperature, and trapping frequencies as the inputs for each data point of Fig. 4, which explains why the theory curve is not smooth in Fig. 4.

---



\end{document}